\documentclass[aps,a4paper,floatfix,nofootinbib]{revtex4}

\usepackage[utf8]{inputenc}

\usepackage{graphicx,color}
\usepackage{amsmath,amssymb,amsfonts,amsthm,amscd,bm}
\usepackage{booktabs}
\usepackage{cancel}

\def\bar{\overline}
\def\hat{\widehat}
\def\*{\star}
\def\[{\left[}
\def\]{\right]}
\def\({\left(}      

\def\){\right)}

\def\frac#1#2{\dfrac{#1}{#2}}
\def\inv#1{\dfrac{1}{#1}}
\def\half{\tfrac{1}{2}}
\def\d{\partial}

\def\rvac{\hbox{$\vert 0\rangle$}}
\def\lvac{\hbox{$\langle 0 \vert $}}
\def\2pi{\hbox{$2\pi i$}}

\def\dsl{\raise.15ex\hbox{/}\kern-.57em\partial}
\def\Dsl{\,\raise.15ex\hbox{/}\mkern-.13.5mu D}
\def\CA{{\cal A}}      
      \def\CF{{\cal F}}

   \def\CN{{\cal N}}   \def\CO{{\cal O}}
      
\def\CS{{\cal S}}      \def\CU{{\cal U}}
\def\CV{{\cal V}}      

\def\rvac{\hbox{$\vert 0\rangle$}}
\def\lvac{\hbox{$\langle 0 \vert $}}

\def\2pi{\hbox{$2\pi i$}}

\def\dsl{\raise.15ex\hbox{/}\kern-.57em\partial}
\def\Dsl{\,\raise.15ex\hbox{/}\mkern-.13.5mu D}
\font\numbers=cmss12
\font\upright=cmu10 scaled\magstep1
\def\stroke{\vrule height8pt width0.4pt depth-0.1pt}
\def\topfleck{\vrule height8pt width0.5pt depth-5.9pt}
\def\botfleck{\vrule height2pt width0.5pt depth0.1pt}
\def\Zmath{\vcenter{\hbox{\numbers\rlap{\rlap{Z}\kern
    0.8pt\topfleck}\kern 2.2pt
    \rlap Z\kern 6pt\botfleck\kern 1pt}}}
\def\Qmath{
    \vcenter{\hbox{\upright\rlap{\rlap{Q}\kern3.8pt\stroke}\phantom{Q}}}}
\def\Nmath{\vcenter{\hbox{\upright\rlap{I}\kern 1.7pt N}}}
\def\Cmath{\vcenter{\hbox{\upright\rlap{\rlap{C}\kern
                   3.8pt\stroke}\phantom{C}}}}
\def\Rmath{\vcenter{\hbox{\upright\rlap{I}\kern 1.7pt R}}}
\def\Z{\ifmmode\Zmath\else$\Zmath$\fi}
\def\Q{\ifmmode\Qmath\else$\Qmath$\fi}
\def\N{\ifmmode\Nmath\else$\Nmath$\fi}
\def\C{\ifmmode\Cmath\else$\Cmath$\fi}
\def\R{\ifmmode\Rmath\else$\Rmath$\fi}
\def\barray{\begin{eqnarray}}
\def\earray{\end{eqnarray}}
\def\beq{\begin{equation}}
\def\eeq{\end{equation}}

\def\kvec{{\bf{k}}}

\def\Li{{\rm Li}}

\def\AA{\leavevmode\setbox0=\hbox{h}
\dimen0=\ht0 \advance\dimen0 by-1ex\rlap{\raise.67\dimen0\hbox{\char'27}}A}

\def\Li{{\rm Li}}

\makeatletter
\def\iddots{\mathinner{\mkern1mu\raise\p@
\vbox{\kern7\p@\hbox{.}}\mkern2mu
\raise4\p@\hbox{.}\mkern2mu\raise7\p@\hbox{.}\mkern1mu}}
\makeatother

\def\Li{{\rm Li}}

\theoremstyle{plain}

\theoremstyle{remark}

\newtheorem{assumption}{Assumption}
\newtheorem{property}{Property}

\def\alphavec{\bm{\alpha}}
\def\phivec{\bm{\phi}}

\def\gcal{\mathfrak{g}}
\def\pcal{\mathfrak{p}}
\def\rank{\mathfrak{r}}

\def\rhovac{{\rho_{\rm vac}}}

\def\SU{{\rm SU} (N+1)}

\def\vol{\CV}

\def\LambdaBare{\Lambda_0}

\def\kcut{{k_c}}

\def\Dplus{{\scriptscriptstyle{D}}}

\def\Z{\mathbb{Z}}
\def\ZInteger{\mathbb{Z}}

\def\rvac{| {\rm vac} \rangle}
\def\lvac{ \langle {\rm vac} |}
\def\rvacbeta{| {\rm vac;\beta} \rangle}
\def\lvacbeta{ \langle {\rm vac;\beta } |}
\def\rhoLambda{\rho_\Lambda}

\def\Ncut{{N_{c}}}

\def\cvac{c_{\rm vac}}
\def\cvaccoef{C}

\def\equalreg{\doteq}

\def\cuv{c_{\rm uv}}

\begin{document}

\title{Thermodynamic  formulation of vacuum energy density  in flat spacetime and  potential  implications for the cosmological constant}
\author{
 Andr\'e  LeClair\footnote{andre.leclair@cornell.edu} 
}
\affiliation{Cornell University, Physics Department, Ithaca, NY 14850,  USA} 

\begin{abstract}

We propose a  thermodynamical definition of the vacuum energy density $\rhovac$,   defined as 
$\lvac T_{\mu\nu} \rvac = - \rhovac \, g_{\mu\nu}$,   in quantum field theory  in flat Minkowski space  in $D$ spacetime dimensions,  which can be computed 
in the limit of high temperature,  namely 
in the limit $\beta = 1/T \to 0$.    It takes the form $\rhovac = {\rm const} \cdot  m^D$ where $m$ is a fundamental mass scale and ${\rm ``const"}$ is a computable constant which can be positive or negative.         Due to modular invariance $\rhovac$ can also be computed in a different  non-thermodynamic channel where one spatial dimension is compactifed on a circle of circumference $\beta$ and we confirm this  modularity for free massive theories for both bosons and fermions for $D=2,3,4$.     We list various  properties of $\rhovac$ that are generally required,  for instance $\rhovac =0$ for conformal field theories,  and 
others,  such as the constraint that $\rhovac$ has opposite signs for free  bosons verses fermions of the same mass,  which is related to  constraints from supersymmetry.         Using the Thermodynamic Bethe Ansatz we compute $\rhovac$ exactly for 2 classes of integrable QFT's in $2D$ and  interpreting some previously known results.      We apply our definition of $\rhovac$ to Lattice QCD data  with two light quarks (up and down) and one additional massive flavor (the strange quark),  and find  it is negative,  $\rhovac \approx  - \( 200 \, {\rm MeV} \)^4$.   
Finally we make some remarks on  the Cosmological Constant Problem since $\rhovac$ is central to any discussion of it.

\end{abstract}

\maketitle
\tableofcontents

\section{Introduction}

This article is mainly concerned  with the vacuum energy density of a quantum field theory (QFT)  in arbitrary spacetime dimension $D$ for flat Minkowski space.   
The vaccum energy density  $\rhovac$ of a QFT
\beq
\label{rhvacdef0}
\langle T_{\mu\nu} \rangle = \lvac T_{\mu\nu} \rvac  =  - \rhovac \, g_{\mu\nu},
\eeq 
is difficult to compute due to the need to drastically regulate ultra-violet divergences even in free QFT's.    
One needs a consistent prescription and this is the main focus of this article.   
 
 To motivate our proposed prescription,  let us consider a  simple harmonic oscillator with quantized energies 
 $E_n = \hbar \omega (n+ \half)$.    If the SHO is completely isolated,  i.e. nothing else exists,   then it is in fact impossible to measure the zero point energy $\hbar \omega/2$ since both classical and quantum mechanics are invariant under arbitrary shifts of the potential $V \to V + v_0$.     To measure the zero point energy one needs to couple it to some environment.    One could imagine the particle is charged and couple  it to photons,  and measure energies of photons emitted in transitions between levels,  such as for the hydrogen atom.      However  a single measurement will only measure energy differences $E_{n_2} - E_{n_1}$ and not the zero point energy since it is independent of the shift by $v_0$.         On the other hand with many such measurements one could ultimately infer the lowest energy level.  The point is one needs to explore all energy levels in order to determine the lowest one.         
 If the particle is not charged,    one can couple it to universal gravity,   however the effects are  much too small to be measured in a laboratory at present.  
 Finally one is led to the idea of coupling the system to a generic heat bath at temperature $T$.    The advantage of this is that one need not be 
 very specific about the interactions with the heat bath.       The Boltzmann distribution certainly depends on the zero point energy, namely it is not invariant under shifts by $v_0$,     and $\rhovac$  can be extracted from various thermodynamic quantities such as specific heats,  etc.  
However the zero point energy is mingled with high energy states and some work is required to extract it. 
   This is the approach followed in this article.

 This thermodynamic approach was proposed for integrable QFT's in D=2 spacetime dimensions in \cite{Mingling}.    That work relied mainly on the exact Thermodynamic Bethe Ansatz (TBA) to calculate the free energy \cite{ZamoTBA,KlassenMelzer,MussardoBook},  and the proper interpretation of the bulk term (see below for the definition).      The TBA here is a relativistic version of Yang-Yang thermodynamics \cite{YangYang}.   In this article we attempt to extend these ideas to QFT's in $D=d+1$ spacetime dimensions. 
 Without the tools of integrability this problem is much more difficult,   nevertheless some general results may be obtained.     
 As we will see,   even  for the massive free field case,  $\rhovac$ has some interesting properties.

In some literature an analogy is made between $\rhovac$ and the Casimir effect.    They are in fact very  different.     The Casimir effect  refers  to the force experienced by conducting plates as a function of their separation $a$ and has been measured.   What is measured is actually simply  related to  just the leading term in the free energy density of photons,  namely $\cuv$ in  equations \eqref{cpowers} where $\beta$ is the spacing between the plates.    
These photons have an equation of state $p = \rho/3$ in $D=4$ spacetime dimensions,   which is not at all the same as $\rhovac$  considered here.  
This is clear from the standard formulas for the leading contribution to the Casimir force which only depend on $\hbar$ and $a$,  and not on the fine structure constant $e^2$,  nor the mass of the electron,    where $\beta$ plays the role of the spacing $a$,  which can be derived from \eqref{cdef}, \eqref{cpowers}.        
In fact we will argue below that $\rhovac =0$ for any conformal field theory (CFT).

 Since $\rhovac$ is central to any discussion of the Cosmological Constant Problem (CCP),   we begin with remarks concerning this in the next section. 
 There we point out that there is some ambiguity in formulating the CCP depending on what the researcher assumes the ultimate resolution will be. 
 We therefore describe one well-defined formulation of the CCP that in principle can be verified or ruled out with currently known physics. 
 In Section III we propose a non-perturbative definition of $\rhovac$ based on quantum statistical mechanics and point out some of it's advantages.
 This leads us to list some required properties of $\rhovac$,   one of which relies on a restricted modular invariance.    In Section IV we calculate 
 $\rhovac$ for massive free QFT's in $D=2,3$ and $4$ spacetime dimensions and perform a check of modularity.    In Section V we return to 
 $2D$ integrable theories.    We first consider the sine-Gordon model and point out how $\rhovac$ is consistent with certain (fractional) supersymmetric points.   Next we consider theories with a spectrum of particles,   the affine Toda field theories,   and discuss the known fact that 
 $\rhovac$ can be obtained from the S-matrix for the {\rm lightest}  particle scattering with itself and depends only on the mass $m_1$ of this lightest particle.  
 We show how to obtain this result at weak coupling in a semi-classical approach,    where $\rhovac \propto m_1^2/\gcal$ where $\gcal$ is a coupling and the free field limit is $\gcal \to 0$.      After performing these consistency checks,  we apply these ideas to QCD with two light and one heavy quark.    Using Lattice QCD data at high temperatures we extract the vacuum energy and find it is negative,  $\rhovac = - (200\, {\rm MeV})^4$.


\section{Remarks on the Cosmological Constant Problem}

Since $\rhovac$ is central to any potential resolution of the Cosmological Constant Problem (CCP),    in this section 
we make some relevant remarks and provide one well-defined approach where our definition of $\rhovac$ below plays the central role.

\subsection{Ambiguities in formulating the CCP}

Perhaps one reason there has been little progress on the CCP is that it depends strongly on the assumptions made concerning its final resolution.  
Different assumptions can lead to very different approaches to the problem.     For a philosophical discussion on this point see
\cite{Philosophy}.        In the next subsection we will limit the scope and properly define a version of the CCP that is closest to 
Weinberg's original formulation \cite{Weinberg} and can potentially be verified or falsified based on currently known high energy physics.

Let us begin with what is actually measured.      The standard model of cosmology is based on Einstein's equations 
\beq
\label{GRCosmo}
R_{\mu\nu} - \half g_{\mu\nu} \, R + \Lambda \, g_{\mu\nu} = \frac{ 8 \pi G}{c^4} \, T_{\mu\nu}
\eeq
for the Friedmann-Lema\^itre -Robertson-Walker metric.   Above $T_{\mu\nu}$ is the classical stress tensor for the existing observable matter,  including Dark Matter,
and radiation only,  and takes the form 
\beq
\label{Trhop}
T_{\mu\nu} = {\rm diag} \( \rho, p, p, p \) 
\eeq
where $\rho$ is the energy density and $p$ the pressure.   This $T_{\mu\nu}$  can in principle be measured in a laboratory in flat Minkowski space  independently of its cosmological implications.  
 Although   $T_{\mu\nu}$  involves $\hbar$ in various equations of state,  such as the energy density of photons,  Einstein's equations and the measurement of $\Lambda$ are purely classical.   The cosmological constant $\Lambda$ can be measured in astrophysics,  and observations show that it is small and positive.   
Converting $\Lambda$ into an energy density
\beq
   \label{rhoLambda}
  \rhoLambda = \frac{\Lambda c^4}{8 \pi G} , 
  \eeq
 very large scale astrophysical measurements give \cite{WMAP}
 \beq
 \label{rhoLambda}
    \rhoLambda \approx 10^{-9}  ~ {\rm Joule}/{\rm meter}^3 \approx  (0.003 \, {\rm eV} )^4 .
    \eeq

  If one views $\Lambda$ as a new  fundamental  constant that is  independent of Newton's constant $G$ in classical General Relativity,   then
  there really isn't a CCP.     Rather the CCP arises when one considers whether the zero point vacuum energy gravitates.  
  To be precise we should then consider the following:
\beq
\label{GRCosmo}
R_{\mu\nu} - \half g_{\mu\nu} \, R + \LambdaBare \, g_{\mu\nu} = \frac{ 8 \pi G}{c^4} \, \( T_{\mu\nu} + \langle T_{\mu\nu} \rangle \),
\eeq
where 
\beq
\label{Tvac}
 \langle T_{\mu\nu} \rangle = \lvac T_{\mu\nu} \rvac  =  - \rhovac \, g_{\mu\nu},   
 \eeq
 and $\LambdaBare$ is the ``bare"  CC.\footnote{   
 In the above we assume the flat space metric has the signature  $g_{\mu\nu} = {\rm diag} ( -1, 1, 1,1)$. }   
Moving the $\LambdaBare$ term to the RHS one obtains the effective 
\beq
\label{rhoeff}
\rhoLambda=  \rhovac +  \rho_{\Lambda_0} 
\eeq
From this point of view,   it is the above $\rhoLambda$ that is measured to be the numerical value in \eqref{rhoLambda}.   

\bigskip

The various versions of the CCP in the literature can be delineated in the following partial list,  whose items are  not necessarily mutually exclusive, and are all  currently reasonable assumptions:       
(i)  It may be that $\LambdaBare$ is a new fundamental constant of Nature  and understanding it is like trying to understand why $G \neq 0$.   ~ 
(ii)  Perhaps $\rhovac$ does not gravitate.    This assumption can be motivated by the fact that classical mechanics is unaffected by a shift of the potential energy,  which shifts the zero point energy.   The same is true for an isolated quantum mechanical system.   In this case 
$\rhoLambda$  only arises from the free parameter $\LambdaBare$ which requires the new fundamental constant $\LambdaBare$.  
On the other hand,  as for $T_{\mu\nu}$,   in principle $\rhovac$ can be measured independently of cosmology and should gravitate. 
~
(iii)  If $\rhovac$ does gravitate,  and if it  does not give the measured value when properly calculated,   then a non-zero   $\LambdaBare$  must be invoked to fine tune to the measured value.   
~
(iv)   The issue can only be resolved by a complete formulation of Quantum Gravity,  namely quantum fields in quantized spacetime,  or at the very least QFT in curved spacetime.   
In particular the assumption is that  $\rhovac$ cannot be  simply computed in flat Minkowski space.    

\bigskip

\subsection{Definition of one formulation of the CCP and its  modest assumptions}

Regardless of one's point of view on what the eventual resolution entails,   $\rhovac$ is central to the CCP.     Let us  now limit the scope of (i)-(iv) above in such a manner that the CCP 
can be  resolved  in principle under these assumptions with our current understanding of QFT in  Minkowski space.     
Certainly one can take issue with these assumptions,   however for further progress on the CCP,     it is a worthwhile exercise to consider this conservative version,  and other well-defined versions,  in order, at the very least to  to understand if they are ruled out or not.   

\bigskip

\begin{assumption} 
{\it A Principle of Nothingness.} ~~ 
We assume the bare $\LambdaBare =0$.     One can argue in favor of this as follows.    Suppose the Universe is {\it void},    which is to say all the known particles do not exist.   One can then wonder whether the Universe exists at all!    In any case,  this  implies  $T_{\mu\nu} =0$  as a quantum operator and thus $\langle T_{\mu\nu} \rangle = 0$.      
In such a void,   Newton's constant doesn't appear in Einstein's equations but only $\Lambda_0$,  and the solution is deSitter space.  Thus Minkowski space is unstable.     This Principle of Nothingness is equivalent to demanding that Minkowski space is stable in the void.   
\end{assumption} 

\bigskip
\begin{assumption}
$\rhovac$ does gravitate.     There is no reason to expect that it doesn't,   since the zero point energy is measurable when coupled to an environment such as a heat bath.   
As we will explain below,   in principle $\rhovac$ can be measured by finite temperature measurements in Minkowski space.  
\end{assumption} 

\bigskip
\begin{assumption}
$\rhovac$ can be computed in Minkowski space without quantum gravity.       In support of this,  let us mention that in the standard model of Cosmology,  
one does not need to understand  QED,  i.e. quantized  radiation (photons) and electrons,  in curved spacetime,  to understand for instance the Cosmic Microwave Background of photons.      
\end{assumption}

\bigskip
\begin{assumption}
A full theory of Quantum Gravity and matter is not necessary to understand the measured $\rhoLambda$.   This is an assumption which may turn out to be wrong.      We make this assumption since a complete theory of Quantum Gravity may be out of reach for a very long time,   so it is worthwhile understanding whether the CCP can be explained by currently known physics,  even  if only to rule out this possibility.      
String Theory is a promising candidate,  but to date cannot predict $\rhovac$;  furthermore positive values of the CC appear to be rare in the space of vacua in String Theory.   It is  however a theory with a single mass scale,  which is essentially the Planck scale,  and the space of string vacua does not have an independent parameter 
$\LambdaBare$,         Let us also mention that measurements of $\rhoLambda$ are on very large length scales where one does not expect quantum gravity to be significant,  in contrast  with the  initial Big Bang singularity or the singular black hole interior.   
\end{assumption}


\def\SC{SpC~}

\section{Non-perturbative definition of vacuum energy based on thermodynamics in any dimension}

\subsection{Two channels and modular invariance}

\begin{figure}[t]
\centering\includegraphics[width=.7\textwidth]{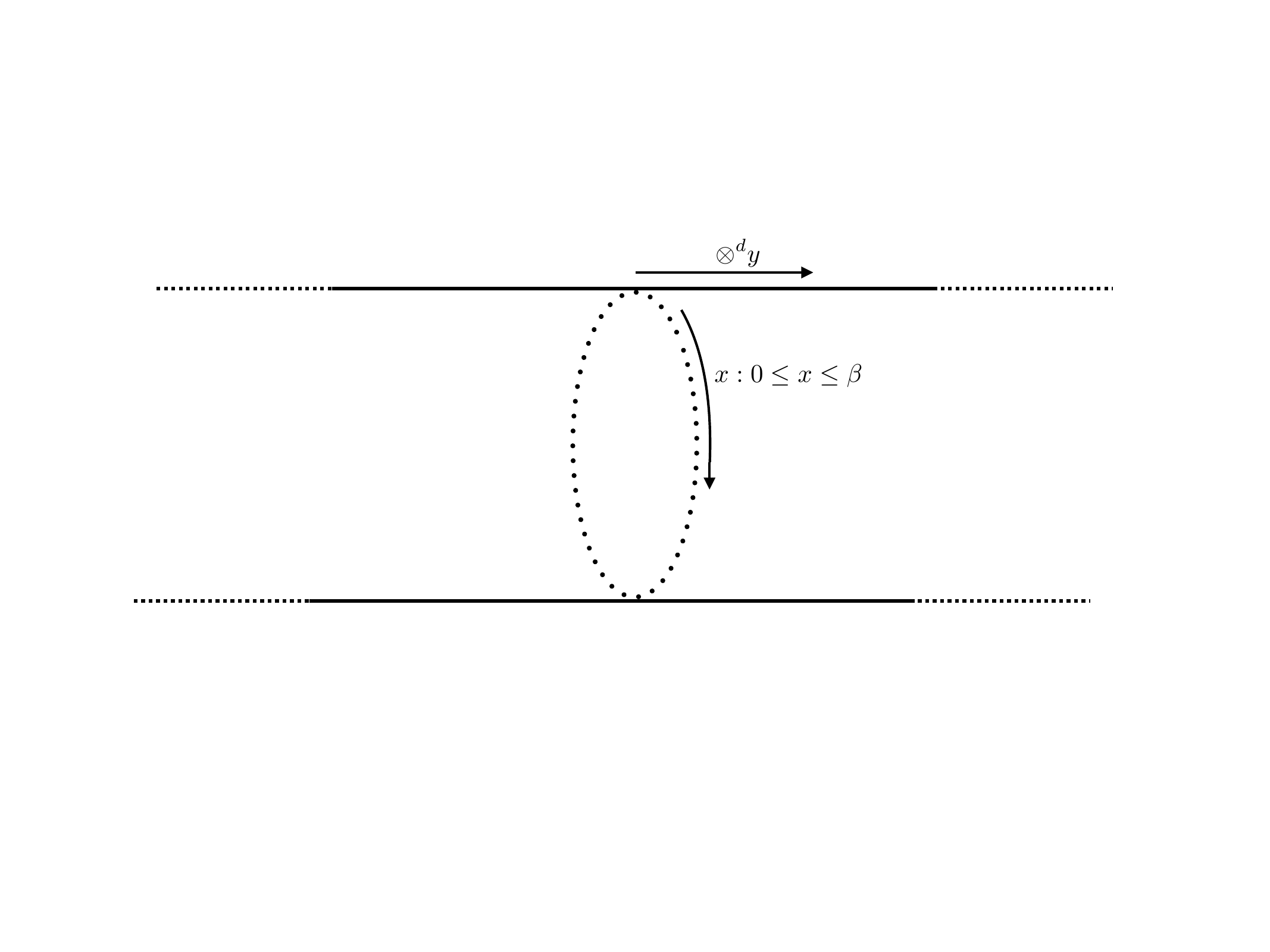}
\caption{An infinite cylinder in $D=d+1$ spacetime dimensions.       For the thermal channel,  $x$ is euclidean time and $y$ refers to $d$ spatial directions.      For the \SC channel $x$ is a single compactified spatial coordinate and one $y$ coordinate is time 
$-\infty < t < \infty$.}
 \label{LRchannel}
\end{figure}

Consider a QFT in euclidean space where one of the dimensions is compactified on a circle of circumference $\beta$.    
See Figure \ref{LRchannel}.
There are two points of view of such a theory.    

\bigskip

\noindent {\bf Thermal channel.} ~~
 If the compactified direction is euclidean time,  then this is interpreted as the 
QFT at finite temperature $T=1/\beta$.    In this picture  the Hilbert space corresponds to particles of arbitrary
continuous momentum $\kvec$ in $d$ spatial dimensions.     We will refer to this as the {\it Thermal}  channel.    

\bigskip
\noindent {\bf Spatially compactified channel.} 
The compactified direction can be viewed as a spatial coordinate.    For $D=2$,   one would say the hamiltonian ``lives"  on the circle,  i.e. the  Hilbert space is the quantized states on a circle,
and the other direction corresponds to time evolution.    In this case,  {\it one} component of the spatial momenta are not continuous but quantized.
We will refer to this as the {\bf \SC}  channel.\footnote{In \cite{LeClairMussardo} these two channels were referred to as $R$ verses $L$ respectively.} 
 
 \bigskip
 As we will see below even for free QFT's,   the Thermal channel has several advantages.  One is that the integrals involved are convergent if the theory is UV complete. 
 On the other hand,  the \SC channel requires  a detailed regularization that in principle makes it non-universal.     As discussed below Modularity can serve as a constraint for the regularization procedure.

 \subsection{Thermodynamic definition of $\rhovac$}
 
 Let us first consider the Thermal channel.    As usual we define the partition function and free energy density $\CF$,
\beq
\label{Z}
Z = {\rm Tr} \, e^{-\beta H}, ~~~~~ \CF = - \inv{\beta \vol}  \log Z
\eeq
where $\vol$ is the $d$  dimensional uncompactified spatial volume.   The energy density and pressure are given by 
\beq
\label{prho}
\rho = \d_\beta \( \beta \CF \), ~~~~~p = - \CF
\eeq
and the entropy density is 
\beq
\label{Entropy}
\CS = \beta^2\,  \frac{\d\CF}{\d\beta}  = \beta\, (\rho + p ). 
\eeq

Let us assume that there is a single fundamental mass scale  $m$ such that all {\it physical} masses are proportional to $m$.  
The Standard Model of particle physics is such a model if one accepts that all fundamental masses arise from spontaneous symmetry breaking in the Higgs sector.    String Theory is also a theory with a single mass scale.    
Then it is convenient to define the scaling function $c(r)$: 
\beq
\label{cdef}
\CF = - \frac{\chi(D)}{\beta^{D}} \, \, c(r),  ~~~~~ r \equiv m\beta ,   ~~~~~ \chi(D) \equiv  \pi^{-D/2} \Gamma (D/2) \zeta (D).
\eeq
The function $\chi (z)$ satisfies the functional equation 
\beq
\label{func}
\chi (z) = \chi(1-z) 
\eeq
in the entire complex plane (except at the pole at $z=1$),    as first discovered by Riemann.    This functional equation will be be crucial to 
Modularity discussed below.   
The normalization in \eqref{cdef} is such that for a free massless boson,   $\lim_{r \to 0} c(r) = \cuv =1$ independent of $r$.
For $D=2$ unitary theories,   $\cuv$ is the Virasoro central charge.  

For unitary theories in 2D,   the c-theorem implies $c(r)$ decreases with $r$\cite{ctheorem}.   
For our thermodynamic definition of $c(r)$, 
\beq
\label{ctheoremeq}
\d_r c(r) < 0 ~~~~~~~ \Longrightarrow ~~\rho > (D-1) p .
\eeq
For  the free massive free theories considered below,   this condition is satisfied.   As we will see below  it is also satisfied for QCD (See Figure \ref{QCDfig}.)
It would be nice to show  that $c(r)$ decreases based on general principles,  however this is beyond the scope of the present article.

\def\cuv{c_{\rm uv}}

If the QFT is UV complete,  then there should be a power series expansion in $r$:
\beq
\label{cpowers}
\lim_{r \to 0} \, c(r) = \cuv + \sum_{\pcal >0}  c_\pcal \, r^\pcal  
\eeq
where $\pcal$ denotes certain powers.   These powers are not necessarily integers,   but can depend on the anomalous scaling dimension of 
relevant operators in the UV.     Also as we will see,   some powers have $\log r$ corrections.

\def\Dsmall{\Dplus}
\def\cvaccoef{\CA}

\bigskip

\def\cvaccoef{c_{\Dplus}}

\noindent{\bf Definition of $\rhovac$.}~~~~
Vacuum energy with $p = - \rho$ would correspond to $\d_\beta \CF =0$ based on \eqref{prho},  and this is not satisfied for arbitrary $\CF (\beta)$ as expected.   However if $c(r) \sim r^D$ then $\d_\beta \CF =0$.     We thus identify $\rhovac$ with the $c_\Dplus$ term:        
\beq
\label{cvac}
\cvac (r) = c_\Dplus  r^D , 
\eeq
where $\cvaccoef$ is a constant.   
This leads to the definition 
\beq
\label{rhovacC}
\rhovac = -  c_{\Dplus}  \, \chi (D)  \,  m^D.
\eeq
Since $r \to 0$ corresponds to $\beta \to 0$ for fixed $m$,   the above $\rhovac$ is effectively calculated in the UV since it corresponds to
the limit of infinite temperature.

Now consider the \SC channel.    Here $\beta$ is not interpreted as a temperature but rather a finite spatial size.    
Formally one has
\beq
\label{sC}
\CF =  \lvacbeta \, H \rvacbeta
\eeq
where $\rvacbeta$ is a complicated $\beta$-dependent vacuum.   As we will see,   computations in this channel are more complicated than
for the Thermal channel since it requires some subtle regularization of divergences.

\def\psibar{\bar{\psi}}

\subsection{Required properties of $\rhovac$}

In this section we list several properties of $\rhovac$ we require to be valid,   as they will serve as non-trivial checks of our proposal.

\bigskip
\begin{property} 
We assume that the QFT is UV complete such that the power series \eqref{cpowers} is well-defined.  
\end{property}

\bigskip
\begin{property} 
If the QFT is conformally invariant,  i.e. a CFT,   then $\rhovac =0$.       The reason is simple:    the CFT only has the leading 
$\cuv$ term in the expansion \eqref{cpowers}.    This implies free massless particles have $\rhovac =0$.    
\end{property} 

\bigskip
\begin{property}
The entropy associated with the vacuum state is zero.     This is evident from \eqref{Entropy} with $p = -\rho$.
\end{property}

\bigskip

\begin{property}
{\bf Bose-Fermi symmetry.}     Introduce a statistical parameter $s$,  where $s=1, -1$ corresponds to bosons, fermions respectively. 
For the complete $\CF (r)$ there is no Bose-Fermi symmetry for a particle of a fixed mass since there is no such symmetry between
the Bose-Einstein  (BE) and Fermi-Dirac (FD) distributions.    However for the term that defines $\rhovac$,  we expect  that  bosons and fermions 
contribute with opposite signs,   namely $\rhovac \propto s$.       The arguments  are  simple and well-known,   nevertheless we repeat them here, in part since we wish to present an analogous result for interacting theories.

\bigskip
Consider  the  action for a single bosonic field  $\phi$ in zero spatial dimensions:
\beq
\label{bose.1}
S = \int dt \( \inv{2}  \d_t \phi \d_t \phi - \frac{\omega^2}{2} \phi^2 \)
\eeq
The equation of motion is 
$\d_t^2 \phi = - \omega^2 \phi$,  thus $\phi $ can be expanded as follows:
$\phi =  \(  a e^{-i\omega t}  + a^\dagger e^{i\omega t} \)/ \sqrt{2\omega}$.
Canonical quantization gives
$[a, a^\dagger] = 1$,  and the hamiltonian is
\beq
\label{bose.3}
H = \inv{2} \(  \d_t\phi \d_t \phi  + \omega^2 \phi^2 \)  =  \frac{\omega}{2} ( a a^\dagger + a^\dagger a)  =    \omega ( a^\dagger a + \inv{2} )  
\eeq
The zero-point energy is identified as $\omega/2$.   

For fermions,  the zero point energy has the opposite sign.  To see this, consider 
a zero dimensional  version of a Majorana fermion with action
\beq
S = \int dt  \(  i \psibar \d_t \psibar - i \psi \d_t \psi - \omega \psibar \psi \)
\eeq
The equation of motion is 
$ \d_t \psi = -i \omega \psibar$, and $ \d_t \psibar = -i \omega \psi $,    
and thus $\psi, \psibar$ have the following expansion:
$\psi  =   \inv{\sqrt{2}} \(  b e^{i \omega t} + b^\dagger e^{-i\omega t}  \)$, and  
$\psibar =  \inv{\sqrt{2}} \( - b e^{i \omega t} + b^\dagger e^{-i\omega t}  \)$. 
Canonical quantization leads to
$\{b, b^\dagger\} = 1,  b^2 = {b^\dagger}^2 =0$,  
and the hamiltonian is 
\beq
H =  \omega \psibar \psi = \frac{\omega}{2} \(  b^\dagger b - b b^\dagger \) =  \omega ( b^\dagger b - \inv{2} )
\eeq
and the zero point energy is $-\omega/2$.   

For non-free theories,  this property also follows trivially from supersymmetry .  It is well-known that a finite temperature breaks supersymmetry,  
for the simple reason that there is no such symmetry between the BE  and FD distributions \cite{Grisaru}.   
However  a simple argument shows that supersymmetry implies that $\rhovac =0$.    Let $Q$ be a conserved supercharge where schematically 
$Q^2 = H$.    If $Q \rvac =0$,  then $\lvac H \rvac =0$,   implying that bosons and fermions have zero point energies of opposite sign, 
and $\rhovac =0$.    

Let us also mention a generalization that is realized in $2D$.      Suppose the hamiltonian $H$ is in  the center of the  universal enveloping algebra 
of some conserved charges $Q$.    For instance the sine-Gordon model at certain rational points of the coupling constant 
has  conserved charges satisfying $Q^p = H$ where $p$ is an even integer \cite{Vafa}.     Then by the same argument,  if
$Q \rvac =0$ then $\rhovac =0$.      We will confirm  this below when we consider $\rhovac$ for the sine-Gordon model.  
\end{property}

\bigskip

\begin{property}
{\bf Modularity.} ~~~Deceptively simple as it may appear,  calculations of $\rhovac$ in the Thermal and \SC channels must agree.   
As we will show below,   this is rather non-trivial since the calculations are completely different.      In these computations, 
Riemann's functional equation \eqref{func} is crucial.   

For a $2D$ conformal field theory,  one can impose boundary conditions on our infinite cylinder such that it is a torus.  
Here the partition function enjoys the full $SL(2, \ZInteger)$ modular invariance \cite{Cardy}.    A subgroup of  the modular group simply exchanges the two cycles, i.e. exchanges space and time,   which is the distinction between the Thermal and \SC channels.     For non-conformal theories in arbitrary dimensions we expect the latter to hold not only for $\rhovac$ but for the entire $\CF$,    and will refer to it as Modularity.  

\end{property}

\bigskip
\begin{property} 
For this article,  we don't implicitly assume nor consider  spontaneous symmetry breaking (SSB).      We will see that in QCD for instance,  one does not need to
assume SSB to compute $\rhovac$.     We should also mention that examples with SSB of supersymmetry were already  considered in \cite{Mingling}. 
\end{property}


\def\Omegad{\Omega_d} 
\def\cvac{c_{\rm vac}}
\def\cvaccoef{C}

\def\rhovaccutoff{\rho_{\rm vac, cutoff}}

\section{The massive  free field case in $D$ spacetime dimensions}

\subsection{The naive computation that leads to a discrepancy by $10^{120}$ orders of magnitude}

Although this computation should be obsolete by now,   we present it here since the discrepancy by a factor of 
$10^{120}$  is still often quoted in the literature.    It will also serve as a point of comparison with a properly regularized definition below.    Consider a free massive  bosonic or fermionic field,   which can be viewed as an infinite collection of harmonic oscillators with zero point energy
$\sqrt{\kvec^2 + m^2}$:
\beq
\label{rhoWein}
\rhovaccutoff = \frac{s}{2} \int_{|\kvec|  \leq \kcut}  \frac{d^d \kvec}{(2 \pi)^d} \, \sqrt{\kvec^2 + m^2 }.
\eeq
There is already one problem in that if one naively computes the pressure using a similar expression,    it does not satisfy $p = -\rho$ \cite{Martin}. 
Using 
\beq
\label{OmegaDef}
\int \frac{d^d \kvec}{ (2 \pi)^d}  = \Omegad \int k^{d-1} dk, ~~~~~~ \Omegad = \inv{\Gamma(\tfrac{d}{2}) 2^{d-1} \pi^{d/2} } ~~~~~~ k = |\kvec|
\eeq
the integral can be performed for arbitrary $d$ and expressed in terms of a hyper-geometric function:  
\beq
\label{hyper}
\rhovaccutoff  =  s \, \(  \frac{ m\, \kcut^d  }{2d} \) \,  \Omega_d  ~_2 F_1 \[ -\half, \tfrac{d}{2} , \tfrac{2+d}{2}, -  \tfrac{\kcut^2}{m^2} \]. 
\eeq

For our purposes it is more useful to consider each integer spacetime dimension $D$ separately.   
For $\kcut \gg m$ one finds 
\beq
\label{rhoWein}
\rhovaccutoff =  s
\begin{cases}
\frac{\kcut^2}{4 \pi} + \frac{m^2}{4 \pi} \,  \log (2 \kcut/m )   ~~~~~~~~~(D=2) \\ 
\\
\frac{\kcut^3}{12 \pi} - \frac{m^3}{12 \pi}  ~~~~~~~~~~~~~~~~~~~~~~~(D=3) \\
\\
\frac{\kcut^4}{16 \pi^2} - \frac{m^4}{32 \pi^2} \, \log ( 2 \kcut /m  ) ~~~~~(D=4)
\end{cases}
\eeq
The $120$ orders of magnitude arises when one chooses the cut-off $\kcut$ to be the Planck scale and compares with 
the measured value \eqref{rhoLambda}.     Note the $\log$ corrections in even dimensions,  which will reappear below with a proper regularization that 
removes the cut-off dependence.     Note also the difference in signs for $D=2$ verses $D=4$.

\def\cuv{c_{\rm uv}}

\subsection{Thermal Channel}

We begin with a standard result from quantum statistical mechanics:
\beq
\label{Free1}
\CF = \frac{s}{\beta} \int  \frac{d^d \kvec}{ (2 \pi)^d}  \, \log \( 1-s \, e^{- \beta \, \omega_\kvec} \), ~~~~~\omega_k = \sqrt{\kvec^2 + m^2}
\eeq
An important advantage of the Thermal channel is that the above integrals are already convergent.    It remains to extract $\cvac(r)$ from it.  
Rescaling $\kvec \to 2 \pi \kvec/\beta$ 
\beq
\label{Free2}
\CF = \frac{s}{\beta^{d+1} }\Omega_d \, \int_0^\infty  d k  k^{d-1}  \log \( 1-s \, e^{- \, \sqrt{k^2 + r^2} }\), ~~~~~
r = m\beta.
\eeq
Expanding out the $\log$,  
the scaling function $c(r)$ can then be expressed
\beq
\label{cBessel}
c(r) =  \frac{r^{\Dplus/2} \pi^{-\Dplus/2} }{2^{(\Dplus-2)/2} \, \chi(\Dplus)} \, 
             \sum_{n=1}^\infty  \frac{s^{n+1} }{n^{\Dplus/2} } \, K_{\tfrac{\Dplus}{2} } (nr)   
  \eeq
  where $K_\nu (z)$ is the modified Bessel function.  
  Using the expansion 
\beq
\label{Kasym}
\lim_{r \to 0}   K_{\tfrac{\Dplus}{2}} (nr ) = r^{-\Dplus/2} \, 2^{(\Dplus -2)/2}\,  n^{-\Dplus/2} \, \Gamma( \tfrac{\Dplus}{2} ) + \CO( r^{\Dplus/2} ),  
\eeq
and  the following property of the Riemann zeta function
\beq
\label{cuvfb}
\zeta(z) = \sum_{n=1}^\infty \inv{n^z} = (1- \frac{1}{2^{z-1}} )^{-1} \, \sum_{n=1}^\infty  \frac{(-1)^{n+1}}{n^z} , ~~~~~\Re (z) > 1 , 
\eeq
one can easily show that the leading term is 
\beq
\label{cUV}
\lim_{r \to 0}  c(r) = \cuv  = 
\begin{cases} 
1,  ~~{~~~~~~~~~~~\rm  for~ bosons}  ~~~ (s=1) \\
1- \inv{2^{\Dplus-1} },  ~~ {\rm  for~ fermions}~ (s=-1) .
\end{cases}
\eeq

We now turn to the small $r$ corrections in order to determine $\rhovac$.    As we will see,  each spacetime  dimension  $D$  should
be  treated separately.

\subsubsection{$D= 2$}

This case was studied to all orders in $r$ by Mussardo in his book \cite{MussardoBook}.    In fact there Modularity was established to all 
orders.\footnote{There is a small discrepancy  between the result stated in \cite{MussardoBook}  and our result below,  and we think this was a relatively minor error.        
Namely Mussardo finds   a linear in $r$ term in $c(r)$ for bosons but not for fermions, 
see  equation (19.11.6).  We think this is due to an improper treatment of the zero mode for bosons,  which is somewhat delicate. } 

For our purposes of computing the $\cvac (r) \propto r^2$ term we carry out the calculation in a somewhat simpler  but similar fashion.  
As for the TBA we consider 
\beq
\label{dcr}
\inv{r}\d_r c (r) = - \inv{\zeta(2)} \sum_{n=1}^\infty s^{n+1} \, K_0 (n r ), 
\eeq
where 
\beq
\label{Koasym}
K_0 (n r) =  - \( \log r + \log n - \log 2 + \gamma_E  \) + \CO (r^2) .
\eeq

One encounters the following sum which we regulate with the zeta function: 
\beq
\label{sumn}
\sum_{n=1}^\infty   s^{n+1}  \equalreg  s\,  \zeta (0) = -s/2,
\eeq
where the symbol $\equalreg$ denotes regularization based on the  $\zeta (z)$ function.    
One also encounters the sum  $\sum_{n=1}^\infty s^{n+1} \, \log n$.
Using 
\beq
\label{sumn3}
\sum_{n=1}^\infty \frac{s^{n+1}}{n^\epsilon} = \sum_{n+1}^\infty  s^{n+1} \, e^{-\epsilon \log n}  , 
\eeq
one regulates this as 
\beq
\label{sumn4}
\sum_{n=1}^\infty s^{n+1} \, \log n~ \equalreg  - s \, \zeta' (0) = \frac{s}{2} \log 2 \pi,
\eeq
where $\zeta'(z) = \d_z \zeta (z)$.   
Integrating over $r$ one obtains 
\beq
\label{cofr}
c(r) = \cuv + \frac{3 r^2 s}{2 \pi^2} \( \log (4 \pi/r) + \half - \gamma_E \) + \CO(r^4) ,
\eeq
where $\gamma_E = 0.5772..$  is the Euler-Mascheroni  constant.  
From equation \eqref{rhovacC} one finally obtains 
\beq
\label{rhovac2D}
\rhovac = -  s \, \frac{m^2}{4 \pi} \(  \log (4\pi/r)  + \half - \gamma_E  \).
\eeq

\def\Li{{\rm Li}}

\subsubsection{$D = 3$}

This  case has some new features in comparison with $2D$.    Begin with 
\beq
\label{3D1}
c (r)  =   \sqrt{\frac{2}{\pi}}   \,\(  \frac{r^{3/2}}{\zeta(3)} \) \sum_{n=1}^\infty \frac{s^{n+1}}{n^{3/2} } \, K_{\tfrac{3}{2}} ( n r) 
\eeq
For half-integers $\nu$,   $K_\nu (z)$ has simple expressions: 
\beq
\label{K3asymp}
K_{\tfrac{3}{2} } (nr) = \sqrt{ \frac{\pi}{2}}  \, e^{-nr} \, r^{-3/2}  \, \( \frac{1+ nr }{n^{3/2}} \), 
\eeq
which leads to the exact formula 
\beq
\label{c3exact}
c(r) = \frac{s}{\zeta (3)} \[ \Li_3 (s\,e^{-r} ) + r\,  \Li_2 (s \, e^{-r} ) \],
\eeq
where $\Li_z  (x) = \sum_{n=1}^\infty \, x^n/n^z $ is a poly-logarithm.    Expanding the poly-log in powers of $r$:
\beq
\label{c3asym}
\lim_{r \to 0} c(r) =  \cuv  + \inv{\zeta(3)} \[ \frac{s\log (1-s)}{2} \,  r^2  + \frac{r^3}{3(1-s) } \]+ \CO (r^4) .
\eeq

Note that there is no $\log r$ term in comparison with $2D$.    However one needs to regularize the singular $s$ dependence 
for $s=1$,  and we do this again by analytic continuation of $\zeta (z)$.  Using \eqref{sumn} 
\beq
\label{sumn1}
\inv{1-s} = \sum_{n=1}^\infty  s^{n+1} \, \equalreg -s/2.
\eeq
Expanding $\log (1-s) $ and using \eqref{sumn} 
\beq
\label{sumn2}
\log (1 -s ) \equalreg  \frac{1}{4} .
\eeq 
Putting all this together one obtains
\beq
\label{3D2}
c(r) = \cuv + \frac{s\,  r^2}{8 \zeta(3)} - \frac{s\, r^3 }{6 \zeta(3)} + \CO(r^4) .
\eeq

Based on \eqref{rhovacC} one then  finds  the simple result 
\beq
\label{rhovac3D}
\rhovac =s \,  \frac{m^3}{12 \pi} .
\eeq
Note that it satisfies the Bose-Fermi  {\it Property} 3. 


\subsubsection{$D=4$}

Here 
\beq 
 c(r) = -\frac{\pi^2 \beta^4}{\zeta(4)}  \, \CF .
 \eeq
The calculation is similar to the $2D$ case.       To simplify it we  just deal with 
$c(r)$ directly  rather than $\d_r c(r)$: 
\beq
\label{c4D}
c (r) =  \frac{45\, r^2}{\pi^4}  \sum_{n=1}^\infty  \frac{s^{n+1}}{n^2} \,  K_2 (n r) 
\eeq
Expanding the Bessel function 
\beq
\label{K2}
K_2 (n r) = \( \frac{2}{n^2 r^2} - \inv{2}  \)  + \frac{r^2 n^2 }{32}  \( 3  - 4 \gamma_E  -  4 \log (nr/2) \)   + \CO(r^4).
\eeq 
Regularizing sums involving the statistical parameter $s$ as above,  one obtains 
\beq
\label{c4D2}
c(r) = \cuv -  \frac{15 r^2}{16 \pi^2} (3 +s)  -  \frac{ 45 \,  s \, r^4}{16 \pi^4} \( \log(4\pi/r)   +  \tfrac{3}{4}\ - \gamma_E \) + \CO(r^6) .
\eeq

This leads to 
\beq
\label{rhovac4D}
\rhovac = s\,  \frac{m^4}{32 \pi^2} \( \log(4\pi/r)   + \tfrac34 -\gamma_E \), 
\eeq
Note that the $r^2$ term  in $c(r)$ is not proportional to $s$ as expected due to the asymmetry between BE and FD distributions.
However the $r^4$ term which determines $\rhovac$  is  indeed  consistent with Bose-Fermi symmetry {\it Property} 3.   

\bigskip
\noindent 
{\bf Remark.} ~~ Note that for all $3$ cases of $D$,   $\rhovac$ has the {\it opposite} sign to the cut-off expressions \eqref{rhoWein},  which is
presumably  due to the subtraction of divergent terms as $\kcut \to \infty$.      It is also interesting that up to this sign,  the $\log$ terms in even dimensions follows simply by replacing the cut-off $\kcut \to 2/\beta$ which is consistent with the fact that $\beta \to 0$ is the UV limit.

\def\CFsum{\CF_{\rm sum}}
\def\CFint{\CF_{\rm int}}

\subsection{Spatially compactified channel: Modularity}  

The \SC channel is essentially the same as for the naive calculation in Section IVA,  however regularized with $\beta$.   
    Calculations in the \SC channel are more difficult for at least two reasons.   First is that UV divergences are severe.   Riemann's  functional identity 
\eqref{func} will play a central role in regulating these divergences.     Secondly,  the statistical parameter $s$ is not simply a parameter as in the Thermal channel,  such that 
bosons and fermions have to be treated separately.    Given these significant differences,   we carry out calculations of $\rhovac$ in this channel as a check of Modularity,  {\it Principle} 4.     This also serves as a check of Bose-Fermi symmetry,  {\it Principle} 3.    

\bigskip

For arbitrary $D=d+1$,  we begin with 
\beq
\label{spactified1}
\CFsum = \frac{s}{2 \beta}   \sum_n \int \frac{d^{d-1} \kvec}{(2 \pi )^{d-1}} ~ \sqrt{\kvec^2 + \( \frac{2\pi n}{\beta} \) ^2 + m^2 }, 
~~~~~~~{\rm where} ~
\begin{cases}  ~~~~~ 
n\in \Z ~{\rm for}  ~ {\rm bosons} ~(s=1 )\\
 n \in \Z + \half  ~{\rm for}  ~ {\rm fermions}  (s=-1) .
 \end{cases}
\eeq
We require that as for the Thermal channel, in the infra-red (IR)  $\lim_{r \to \infty} \CF (r) \to 0$.    More generally this is a consequence  of  a massive theory where the infra-red limit of the 
QFT is empty since all masses are going to infinity in this limit,   i.e.   $\lim_{r \to \infty} c(r)  =0$.     This also implies the entropy density $\CS$ in \eqref{Entropy}  is zero in this limit,  as it should be.     It is implicit in this condition that the theory does not  have  a  non-trivial IR fixed point CFT, which is of course known to be the case for free  massive QFT's.\footnote{Theories with non-trivial IR fixed points involve massless Goldstone bosons.   Theories with SSB of supersymmetry were also considered in \cite{Mingling},   and these  Goldstone bosons do contribute to $\rhovac$.}

 Following Mussardo \cite{MussardoBook},  we deal with this with the subtraction:
\barray
\label{spactified1}
\CF &= &\CFsum - \CFint \\
\CFint  &=& \frac{s}{2 \beta}   \int_{-\infty}^\infty  dx  \int \frac{d^{d-1} \kvec}{(2 \pi )^{d-1}}  \sqrt{\kvec^2 + \( \frac{2\pi x}{\beta} \) ^2 + m^2 }.
\earray
The above expression for $\CF$  still has UV divergences,  however as we will see,  they are more  easily interpreted.   

\bigskip
We present calculations for $D=2$ bosons and fermions in more detail since similar details appear for 
$D=4$.    We leave the general case of $D=3$ bosons and fermions,   and $D=4$ fermions as an exercise.

\bigskip

\subsubsection{D=2 Bosons}

Here the single  spatial momentum $\kvec$ is quantized and there are no leftover spatial integrals:
\beq
\label{D2Boson}
\CFsum = \inv{2 \beta} \sum_{n \in \ZInteger } \sqrt{m^2 + \( \frac{2 \pi n}{\beta} \)^2 }
\eeq

Expanding out the square root in powers of $r = m \beta$,
\beq
\label{D2Boson2}
\CFsum = \frac{ \pi }{\beta^2} \sum_{n \in \ZInteger } \sqrt{n^2 +  \frac{r^2}{4 \pi^2 } } = 
\frac{ \pi }{\beta^2} \sum_{n \in \ZInteger}  \( |n| + \frac{r^2}{8 \pi^2 \, |n| } \)+ \CO(r^4)
\eeq

For $\zeta (z)$ with $z \neq 1$,  one can rely on analytic continuation:
\beq
\label{sums1}
\sum_{n \in \ZInteger } |n| = 2\, \zeta (-1) =  - \tfrac{1}{6} .
\eeq
On the other hand,   the other sum cannot be dealt with this way because of the pole in $\zeta (z) $ at $z=1$.   
Indeed  one encounters a singularity due to the zero mode $n=0$:
\beq
\label{sums2} 
\sum_{n \in \ZInteger } \inv{ |n| } = 2\, \zeta (1) + \tfrac{1}{0} .
\eeq
We discard this $\tfrac10$ singularity,   since as we will see this is needed to satisfy the Bose-Fermi  {\it Property 3},  as there is no zero mode for fermions. 
Leaving the singular $\zeta(1)$ as unevaluated for the moment,  in  summary we have 
\beq
\label{D2Boson3}
\CFsum = - \frac{\pi}{12 \beta^2} + \frac{r^2}{4 \pi \beta^2} \, \zeta (1) + \CO (r^4) .
\eeq

Next we turn to $\CFint$.    The integral over $x$ is still divergent,   thus we 
 introduce a cut-off $\Ncut$ in the integral over $x$:
\beq
\label{D2Boson4}
\CFint =  - \frac{2 \pi}{\beta^2} \, \int_0^\Ncut dx \, \sqrt{x^2  +  (r/2\pi)^2 }.
\eeq
Expanding out the square-root,  
\beq
\label{D2Boson5} 
\CFint = \frac{\pi}{\beta^2} \( \Ncut^2 + \frac{r^2}{4 \pi^2} \( \log(4\pi/r ) + \log \Ncut + \half \)  \) + \ldots 
\eeq

We now regularize $\zeta (1)$ in \eqref{D2Boson3} in a way that is equivalent to the $N_c$ regularization of $\CFint$.  
Using the  harmonic number:
\beq
\label{zetareg}
\zeta_c (1) = \sum_{n=1}^\Ncut \inv{n} = \log \Ncut + \gamma_E + \ldots 
\eeq
Putting this all together,  the $\log N_c$ cancels and  one reproduces  equations  \eqref{cofr} and \eqref{rhovac2D} for $s=1$.

\subsubsection{D=2 fermions} 

Here the calculation nearly identical,   except for the sums over $n$ half-integer.      This implies there is no zero mode $n=0$ to deal with. 
The sums over $n$ can be dealt with using the Hurwitz $\zeta$ function
\beq
\label{Hurwitz}
\zeta (z, a) = \sum_{n=0}^\infty  \inv{(n+a)^z} ,
\eeq
which satisfies
\beq
\label{Hurwitz2} 
\zeta(z,\half) = (2^z -1) \zeta (z).
\eeq
This leads to 
\barray
\label{fermion1}
\sum_{n \in \ZInteger  + \half} | n| &=& 2 \,\zeta(-1,\half) = \tfrac{1}{12} \\
\sum_{n \in \ZInteger  + \half} \inv{|n|} &=&  2 \,\zeta(1,\half) = 2\, \zeta (1) .
\earray
Repeating arguments for the  $2D$ boson case in the last sub-section,   the final result again agrees with equations  \eqref{cofr},\eqref{rhovac2D},  in this case for $s=-1$.

\subsubsection{D=4 bosons} 

In this case there are two one leftover spatial $\kvec$ integrals.    Again rescaling $\kvec \to 2 \pi \kvec/\beta$ one obtains 
\beq
\label{4Da}
\CFsum = \frac{2 \pi^2}{\beta^4}  \sum_{n \in \ZInteger}  \int_0^\kcut  dk \, k \, \sqrt{n^2 + k^2 + \frac{r^2}{4 \pi^2} }.
\eeq
The integral over $k$  is simpler if one considers $\d_r \CF$.   Dropping all $k_c$ dependence as $\kcut \to \infty$ one has 
\beq 
\label{4Db}
\d_r \CFsum = - \frac{r}{2 \beta^4} \sum_{n \in \ZInteger}  \sqrt{n^2 + \( \frac{r}{2 \pi} \)^2 } .
\eeq
Expanding the square-root,
\beq
\label{4Dc}
\d_r \CFsum = - \frac{2 \pi}{\beta^4} \( \frac{r}{2 \pi} \, \zeta (-1) +  \frac{r^3}{16 \pi^3} \zeta(1) \) + \CO(r^3),  
\eeq
and integrating this equation one finds 
\beq
\label{4Dd}
\CFsum (r) - \CFsum(0) =  -\inv{2 \beta^4} 
\( r^2 \zeta (-1) + \frac{r^4}{16 \pi^2} \zeta (1) \) + \CO(r^6).
\eeq
The leading $r=0$ term is 
\beq
\label{4De}
\CFsum(r=0) = - \frac{4 \pi^2}{3 \beta^4} \zeta (-3)
\eeq
which leads to $\cuv = 4 \pi^4 \zeta(-3) /3 \zeta(4)  = 1$, as expected.    

Next consider $\CFint$.   Using essentially the same arguments as for $D=2$,   one finds 
\beq
\label{4Df}
\d_r \CFint = - \frac{r^3}{8 \pi^2 \beta^4} \(  \log \(4 \pi/r \) + \log \Ncut + \half \).
\eeq
Integrating this equation:
\beq
\label{4Dg}
 \CFint = -\frac{r^4}{32 \pi^2 \beta^4}  \( \log \(4 \pi/r  \) + \log \Ncut + \tfrac{3}{4} \)  + \ldots 
 \eeq
  One can easily check that $\CF = \CFsum - \CFint$ gives the same result as equations 
 \eqref{c4D2}, \eqref{rhovac4D} for $s=1$,   hence establishing modularity in this $D=4$ bosonic case.

\section{Integrable theories in $D=2$}

\def\betahat{\hat{\beta}}
\def\GammaUV{\Gamma_{rm uv}}
\def\cpert{c_{\rm pert}}
\def\cbulk{c_{\rm bulk}}

There are many integrable QFT's in $D=2$ spacetime dimensions  available for exploring the above ideas,     and can be useful for 
further developing them.      These integrable QFT's are characterized by a factorizable S-matrix,    and the exact free energy density,  and thus
$c(r)$,  can be computed exactly in the Thermal channel from the TBA,  which involves  a complicated integral equation.     
 The term $\cvac(r)  \propto r^2$  which determines $\rhovac$ is commonly referred to as the
``bulk" term $c_{\rm bulk}$.    For diagonal scattering,  namely where scattering of particles of type $a$ and $b$ produces particles of the same type,
there exists a universal formula for $\cvac$ due to Al.  Zamolodchikov \cite{ZamoTBA}.    
Very interestingly,    $\rhovac$  depends only on physical S-matrix parameters of the scattering of the {\it lightest} particle with itself and the scale of $\rhovac$ is set by the physical mass of this lightest particle.   Since the S-matrices can be bootstrapped starting with the lightest particle,  we henceforth refer to this property as the Lightest Mass Bootstrap (LMB)  property.      Destri and deVega obtained the same result  in the Toda example  below in a very  different manner that is closer to the \SC channel \cite{DestriDeVega};    we will return to discussing the latter work  below.

One subtlety must be kept in mind while interpreting the results.  In 2D the notion of whether a physical particle is a boson or fermion is  blurred  since there is no spin-statistics theorem.     In particular,   the TBA equations are necessarily of fermionic type,   otherwise the solutions are not well-defined in the UV \cite{MussardoSimon}.     In the S-matrix description,  the statistical parameter $s = \pm 1$ is defined as 
\beq
\label{Ss}
\lim_{\theta \to 0} S(\theta) = s 
\eeq
where $\theta$ is the rapidity,    $E= m \cosh \theta, ~ p = m \sinh \theta$,  and for all physical theories with non-trivial S-matrix,  $s=-1$.  

We first consider the sine-Gordon model which has a rich spectrum of particles.    We considered this case already in \cite{Mingling},  
however we repeat some of this here in order to make some additional observations.    As we will see,   $\rhovac$ satisfies Property 3 at certain
(fractional) supersymmetric points.    It will also be interesting to observe how $\rhovac$ behaves near the marginal point.    

The second example is a model of multiple scalar fields,  the affine Toda theories,   where each field leads to a physical particle of a given mass.    
We will show how the exact result at weak coupling can be derived in a semi-classical analysis.    
This example may serve as a tool to understand $\rhovac$ using ordinary perturbation theory in a way that  may lead to insights in higher dimensions,  however this is not pursued here.

\subsection{Exact formula for $\rhovac$}

Let us present the general formula for $\rhovac$ for theories with diagonal scattering  \cite{ZamoTBA,KlassenMelzer}
and explain the LMB property in this context.        
    The basic building blocks of  the two-body S-matrices $S(\theta)$ 
are factors of $f_\alpha (\theta)$: 
\beq
\label{Sf}
S (\theta) = \prod_{\alpha \in \CA}  f_\alpha (\theta), ~~~~~
f_\alpha (\theta) \equiv  \frac{ \sinh \half \( \theta + i \pi \alpha \)}{\sinh \half \( \theta - i \pi \alpha \)} ,
\eeq
and $\CA$ is a finite set of   $\alpha$'s.  
The angles $\pi \alpha$ correspond to resonance  poles in the complex $\theta$ plane for the S-matrices. 

Suppose the theory consists of many particles with masses $m_a, ~ {a=1,2, \ldots}$,  where   $m_1$ is the lightest particle,
and the S-matrix for the scattering of $m_1$ with itself  is $S_{11} (\theta)$ of the form \eqref{Sf}.
Then one has the simple formula  
\beq
\label{cbulkb}
\cvac = - \frac{ 3  r^2}{\pi \gcal}, ~~~~~~ r = m_1 \beta, ~~~~~
\gcal =2  \sum_{\alpha \in \CA}  \sin \pi \alpha .
\eeq
Finally  \eqref{rhovacC} gives\footnote{We have changed the convention for the sign of $\rhovac$ compared to that in \cite{Mingling} in
order to be consistent with higher $D$ conventions above.} 
\beq
\label{cbulkb}
\rhovac  =\frac{ m_1^2 }{2 \gcal} .
\eeq

\def\GammaUV{\Gamma_{\rm uv}}

\subsection{sine-Gordon/massive Thirring  model}

The sine-Gordon theory can be  defined by the following action for a single scalar field:
\beq
\label{shGaction}
\CS = \int d^2 x \( \inv{8 \pi} (\d_\mu \phi \, \d^\mu \phi ) + 2 \mu \cos  ( \sqrt{2} \, \betahat  \phi ) \) .
\eeq
The cosine  term has scaling  dimension $\GammaUV = 2 \betahat^2$,   thus it is relevant for $0 < \betahat^2 < 1$. 
This unitary theory has a UV completion which is just a free boson with $\cuv =1$.   
 In is equivalent to the massive Thirring model with action \cite{Coleman}
 \beq
 \label{Thirring}
 \CS = \int d^2 x \( i \psibar \gamma^\mu \d_\mu \psi - m \psibar \psi - \tfrac{g}{2} (\psibar \gamma^\mu \psi)^2 \), ~~~~~~
 g = \pi \(1/{(2 \betahat^2)} - 1\) .
 \eeq
 The charged solitons of the sine-Gordon model correspond to the Thirring fermion.    The free fermion point occurs at $\betahat^2 = 1/2$.
 For $\betahat^2 < 1/2$  there is a rich spectrum of bound states of the fermion
 and the complete S-matrix was found in \cite{ZamoZamo0}, with the help of the Yang-Baxter equation.   
 
  For generic $\betahat$,    the scattering is non-diagonal.    However
at the couplings 
$\betahat^2 = 1/j,   ~~~~{\rm for} ~~  j=2,3, \ldots$
 then there are $j-2$ breathers and the scattering is diagonal.   These points are referred to as ``reflectionless" due to the diagonal scattering. 
    The soliton is obviously the lightest mass particle.      Letting $s, \bar{s}$ denote the soliton,  anti-soliton,    it's S-matrix is well-known \cite{ZamoZamo0}:
\beq
\label{reflectionlessS}
S_{s \bar{s}} = \prod_{k=1}^{j-2} f_{k/(j-1)} (\theta) .
\eeq
This gives $\gcal = 2 \sum_{k=1}^{j-2} \sin ( \pi k/(j-1) ) = 2 \cot ( \pi/(2j -2) )$.    Expressing $j$ in terms of $\betahat$,   
 then at the reflectionless points 
\beq
\label{rhovacSineG}
\rhovac =   \frac{m_s^2}{4} \, \tan \( \frac{ \pi \betahat^2}{2 ( 1- \betahat^2 )} \) .
\eeq
It was argued that we can analytically continue  the  above formula to $\betahat^2 > 1/2$,  where there are only solitons  in the spectrum. 

\begin{figure}[t]
\centering\includegraphics[width=.5\textwidth]{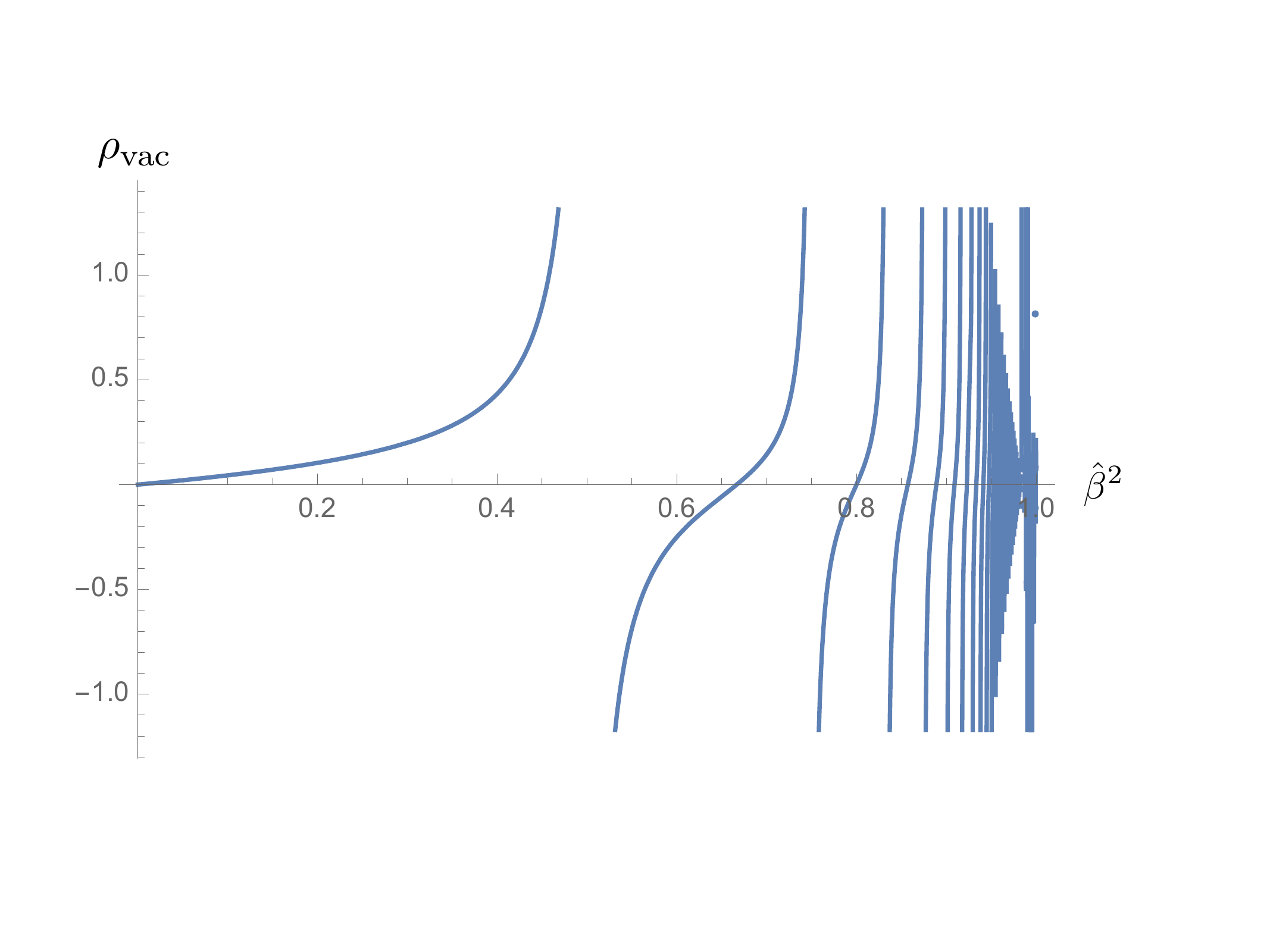}
\caption{$\rhovac$ for the sine-Gordon model as a function of the coupling $\betahat^2$.}
 \label{SGoscillations}
\end{figure}

\bigskip
The above formula has several  very interesting  and illustrative features.    See Figure \ref{SGoscillations}.  

\def\nobull{\bigskip\noindent $\bullet$ ~~}

\nobull
For generic irrational $0< \betahat^2 < 1$,   $\rhovac$ is finite.    This suggests that the $\log r$ corrections in the free theory found in 
\eqref{rhovac2D}  can be eliminated in the interacting theory,   presumably by absorbing them into the definition of physical masses.

\nobull
Below the free fermion point $\betahat^2 = 1/2$,   $\rhovac$ is positive.    At the free fermion point it diverges.     We interpret this reflecting the 
$\log m\beta $ correction in \eqref{rhovac2D} which diverges as $\beta \to 0$.

\nobull
In the limit of weak coupling $\betahat^2 \to 0$,   $\rhovac$ vanishes, in spite of there being an infinite number of bound states in addition to the 
Thirring fermion: 
\beq
\label{betahatlow}
\lim_{\betahat \to 0}  \rhovac =  m_s^2 \( 
\frac{ \pi \betahat^2}{8} + \frac{\pi \betahat^4}{8} + \frac{( \pi^3 + 12 \pi) \betahat^6 }{96} + \CO(\betahat^8 )  \) .
\eeq
It would be interesting to understand this with Feynman diagrams in perturbation theory.  


\nobull
At the marginal point $\betahat^2 = 1$,   $\rhovac$ is indeterminate.     In fact between the free fermion point and the marginal one,   where  the spectrum consists only of the Thirring fermion,   $\rhovac$ undergoes an infinite number of oscillations around zero.   

\nobull
The above $\rhovac$ is consistent with {\it Property} 3 in its supersymmetric version.   Although the notion of a fermion verses boson is ambiguous in 
$2D$,    supersymmetry is not ambiguous.    For $\betahat^2 = 2/3$ the sine-Gordon model has a hidden 
$\CN =2$ supersymmetry \cite{BLfractional},   and indeed $\rhovac = 0$ at this coupling.       However this explains only one of the points where 
$\rhovac =0$.    The other such points also have an explanation.  
In \cite{BernardLeClair} it was shown that the sine-Gordon theory has non-local conserved charges $Q_\pm, \bar{Q}_\pm$  that satisfy the quantum affine algebra
$\CU_q (\hat{{\rm su} (2)}) $ where $q= e^{- i \pi/\betahat^2}$.    When
\beq
\label{betap}
 \betahat^2 = p/(p+1),  
 \eeq
 where $p \geq 1$ is an integer,  $q= - e^{-i \pi /p}$ is a root of unity.    In \cite{Vafa} it was shown that when 
  $p$ is an {\it even} positive integer,   the quantum affine algebra corresponds to a fractional supersymmetry,  
where the conserved charges have spin $\pm 1/p$.      For general $p$,   ``fractional supersymmetry" refers to the fact that mathematically the hamiltonian $H$  is in the center of the  universal enveloping quantum affine algebra,  namely $[Q, H]=0$ for all the non-local charges 
$Q$.    For instance,   for $p=4$,    $Q^4 = H$,   where $Q^4 = Q_+ Q_- Q_+ Q_- + ....$ is quartic  in the conserved charges $Q_\pm$. 
If $Q \rvac = 0$,  then $\rhovac$ should be zero,  and indeed it is for these values of $\betahat$.

\nobull
Finally let us turn to the divergences  in \eqref{rhovacSineG}.   Namely   
   $\rhovac$ diverges at the {\it odd}  $p$  points in  \eqref{betap},  and this is more difficult to explain.     For $p=1$,   above we attributed this divergence to the $\log m\beta$ term in the free fermion theory,  however for higher $p$ this explanation  does not apply since the theory is not  free.
 In \cite{Mingling} we proposed that these divergences arise from a kind of ``resonance" phenomenon.     Namely the terms in \eqref{cpowers} 
with $c_{\pcal > D}$  generally arise from conformal perturbation theory about the UV.     It can happen that these perturbative terms can mix  with
the $\rhovac$ term  
$c_{\Dplus} r^D$ if the anomalous dimension of operators that perturb the  UV CFT  are rational,  such that  higher terms can give an additional  power $\pcal =D$ in \eqref{cpowers}.      For the sine-Gordon model this
  this resonance  condition is simply $p$ an odd integer in \eqref{betap} \cite{Mingling}.    However a better understanding of this divergence is desirable,  at least in order to understand whether it can happen in higher dimensions.   We suspect this is unlikely since anomalous dimensions are not rational in general for $D>2$.

\subsection{ Affine Toda field theories}

In this section we consider  a  model of multiple interacting scalar fields with exponential interactions,  namely the affine Toda theories based on 
SU(N+1).     The S-matrices were first proposed in \cite{Fateev}.  For a comprehensive treatment and review of such theories see \cite{Braden},  and references therein.           In 2D such a theory is super-renormalizable.    We present it to illustrate the LMB  mechanism in this case where each particle corresponds to a scalar field in the theory.  
It is also subject to a weak coupling expansion,  which can be compared with the exact result.   We will carry this out to lowest order in a 
semi-classical approximation.   We will also provide some evidence for a phase transition at the self-dual point.

\subsubsection{Exact $\rhovac$}

Let $G$ denote a finite dimensional Lie algebra of ADE type,  i.e. simply laced,  of rank $\rank$.          Denote the simple roots as $\alphavec_i$,  $i = 1, 2,  \dots \rank$:
\beq
\label{roots}
 \alphavec_i  =   (\alpha_i^1,  \alpha_i^2,  \ldots, \alpha_i^r ) \equiv    \{ \alpha_i^a, ~ a= 1,2, \ldots \rank \}     
 \eeq
 Introduce $\rank$ real scalar fields $\phivec$:
 \beq
 \label{fields}
 \phivec = (\phi^1, \phi^2, \ldots, \phi^r) \equiv \{\phi^a, ~ a=1, 2, \ldots, \rank \}
 \eeq
 such that $\alphavec_i \cdot \phivec = \sum_{a=1}^\rank \alpha_i^a \phi^a$.  
 The untwisted affine Lie algebra $\hat{G}$  has one additional root $\alphavec_0$:
 $ \alphavec_0 = - \sum_{i=1}^\rank  \alphavec_i $. 
 For simply laced $G$ one can take $\alphavec_i^2 = 2  ~~ \forall i$.   
 With these definitions one can define the $2d$ Euclidean action  
 \beq
 \label{AffineAction}
 \CS = \int d^2 x \( \inv{8 \pi} \d \phivec \cdot \d \phivec  +  V(\phivec) \) , ~~~~~
 V (\phivec) = \mu \, \sum_{i=0}^r  e^{b \, \alphavec_i \cdot \phivec } .
 \eeq
 The $1/8 \pi$ normalization corresponds to standard $2d$ conformal field theory conventions 
 where $\langle \phi^a (x) \phi^b (0) \rangle = - \delta_{ab} \log x^2 $ which fixes the convention for the coupling $b$. 
 In particular in the QFT,   $V(\phivec)$ is a strongly relevant perturbation of anomalous scaling dimension $-2 b^2$.

 Henceforth for concreteness we specialize to $G =A_N = \SU$ where $\rank = N$.   Then the Cartan matrix for $\SU$  is $C_{ij} = \alphavec_i \cdot \alphavec_j$,  where the non-zero values for $i \neq j$ are $C_{ij} = -1$ for $i = j \pm 1$ 
 ($ i,j \in \{1,2, \ldots, N \}$).       From the classical mass matrix based on the quadratic term in the potential, 
 \beq
\label{Massterm}
V(\phivec)  = \ldots +  \half \[ M^2 \]_{ab} \phi^a \phi^b + \ldots 
\eeq
one finds 
\beq
\label{MassMatrix}
\[M^2\]_{ab} = 4 \pi \mu\, b^2 \( \sum_{i=0}^N \alpha_i^a \alpha_i^b \). 
\eeq
It is diagonal  
$M^2 = {\rm diag} (m_1^2 , m_2^2 , \ldots m_N^2)$, 
with masses 
\beq
\label{masses}
m_a^2 = 16 \pi \mu \,b^2 \, \sin^2 \( \frac{a \pi}{N+1} \). 
\eeq

Let $S_{11} (\theta)$ denote the S-matrix for the particle of mass $m_1$ with itself:
\beq
\label{S11}
S_{11} = \prod_{\alpha} f_\alpha =  f_{\tfrac{2}{h}} \, f_{- \tfrac{2 \gamma}{h}} \,  f_{\tfrac{2(\gamma -1)}{h}},  ~~~ h = N+1,  ~~~~~~ \gamma = \frac{b^2}{1+b^2}
\eeq 
$h$ is the dual Coxeter number, and  $f_\alpha (\theta)$ is defined above.  
Then
\beq
\label{Toda1}
\gcal = 2 \sum_\alpha \sin \pi \alpha  = 
 - 8 \sin(\pi/h )  \sin (\pi \gamma/h ) \sin (\pi(1-\gamma)/h) . 
\eeq
Based on equation \eqref{rhovacC},   
one then finds
\beq
\label{gcalToda2}
\rhovac =  - \frac{m_1^2}{16 \, \sin (\pi/h ) \, \sin ( \pi \gamma/h ) \, \sin ( \pi (1-\gamma)/h )}.
\eeq

There are some interesting features here also. 
In the zero coupling limit 
this leads to 
\beq
\label{semi}
\lim_{b \to 0} \, \rhovac = 
 - \frac{(N+1)\, m_1^2}{16 \pi \, b^2 \,\sin^2 \( \tfrac{\pi}{N+1} \)} + \CO(1) .
\eeq
An interesting feature is that if the physical mass $m_1$ is fixed then because of the $1/b^2$ dependence it appears non-perturbative.   
However this may simply be a result of expressing in terms of $m_1$ since $m_1 \propto b^2$ in \eqref{masses}.  
Expanding $\rhovac$ for $N=1$ (the sinh-Gordon model) one finds 
\beq
\label{shGasym}
\lim_{b\to 0} \, \rhovac = - m_1^2 \(  \inv{8 \pi b^2} - \inv{8 \pi} - \frac{\pi b^2}{48} + \frac{\pi b^4}{48} - \frac{(60 + 7 \pi^2) \pi b^6}{2880} + \ldots \)~~~~~(N=1) 
\eeq
Again it would be interesting to understand this expansion in perturbation theory.   As previously stated,  the leading $1/b^2$ term may appear
non-perturbative, however  in terms of $\mu$ instead of $m_1 \propto b^2$  it is not.    
Remarkably,   this was  already   understood using the S-matrix bootstrap and  Feynman diagrams by Destri-deVega \cite{DestriDeVega}.
They 
calculated all tadpole diagrams and showed that all the UV divergences could be absorbed into physical masses.

\subsubsection{Semi-classical analysis}. 

Let us show that weak coupling limit in \eqref{semi}  is in perfect agreement with a semi-classical analysis.  
The vacuum energy in this limit is simply the minimum of the potential:
\beq
\label{saddle}
\frac{\delta V}{\delta \phivec} \Bigl|_{\phivec = \phivec_0} = 0
\eeq
which  implies $\phivec_0 = 0$.   
Thus 
\beq
\label{rhovacClass}
\lim_{b \to 0} \,\rhovac = V(\phivec_0) = (N+1) \mu.    
\eeq
Expressing $\mu$ in terms of $m_1$ as given in \eqref{masses} one obtains precisely 
\eqref{semi}, {\it apart from the sign}.       We attribute  this sign discrepancy to the fact that in the thermal TBA channel in 2D, the particles must be treated as fermions rather than bosons.    The higher order terms in \eqref{shGasym} must arise from the corrections to this semi-classical approximation.

\subsubsection{Evidence for a phase transition at $b=1$.}

\begin{figure}[t]
\centering\includegraphics[width=.4\textwidth]{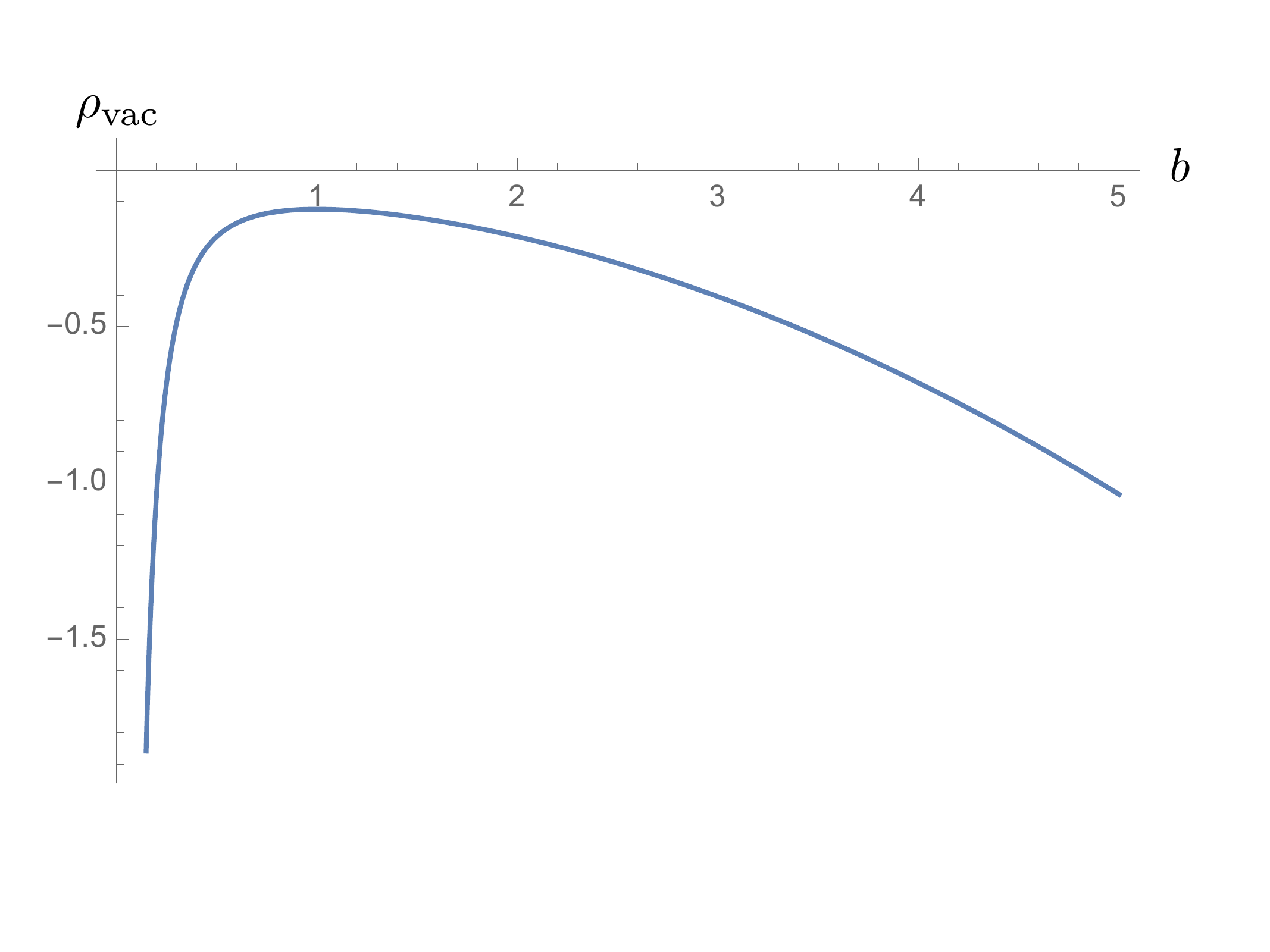}
\caption{$\rhovac$ for the sinh-Gordon model  (N=1) as a function of $b$ based on equation \eqref{gcalToda2}.}
 \label{Todavac}
\end{figure}

It turns out $\rhovac$ is always {\it negative}.      The case $N=1$ corresponds to the sinh-Gordon model.    A plot of $\rhovac$ in this case is 
shown in Figure \ref{Todavac},  and is very similar for other $N$.       
More importantly,    it suggests a quantum phase transition in the vicinity of $b=1$,   since increasing $b$ starting from zero $\rhovac$ increases until
$b=1$ then it starts to decrease.   
  This property relates to a long-term open question concerning the 
properties of the sinh-Gordon above the self-dual point of the S-matrix at $b=1$.     The same issue arises for all the affine Toda theories.      The S-matrices  and $\rhovac$ satisfy the strong-weak coupling duality 
$b \to 1/b$,   however the action has no such symmetry.  Thus  the definition of the sinh-Gordon theory by analytic continuation from  the perturbative region $0<b<1$ to $1<b < \infty$ using the duality is questionable.    This has been studied in great detail using various methods in \cite{Konik},   and significant evidence for some phase transition at $b=1$ is seen,  however a description of the theory for $b>1$ was beyond their  scope,   except that it was proposed there that the theory is massless.       Recently a proposal for the sinh-Gordon beyond the self-dual point was proposed in \cite{BLfreezing} by invoking a
Coulomb background that is shifted from the Liouville one,   and this reproduces some known results in disordered systems;     however  a complete description of the theory above $b=1$  is still lacking.  
 Nevertheless it is interesting that $\rhovac$ gives some evidence for such a phase transition if our interpretation is correct.    

\section{QCD in 4D: ~~ Estimation of $\rhovac$ from numerical lattice results for 3 flavors.}

\begin{figure}[t]
\centering\includegraphics[width=.5\textwidth]{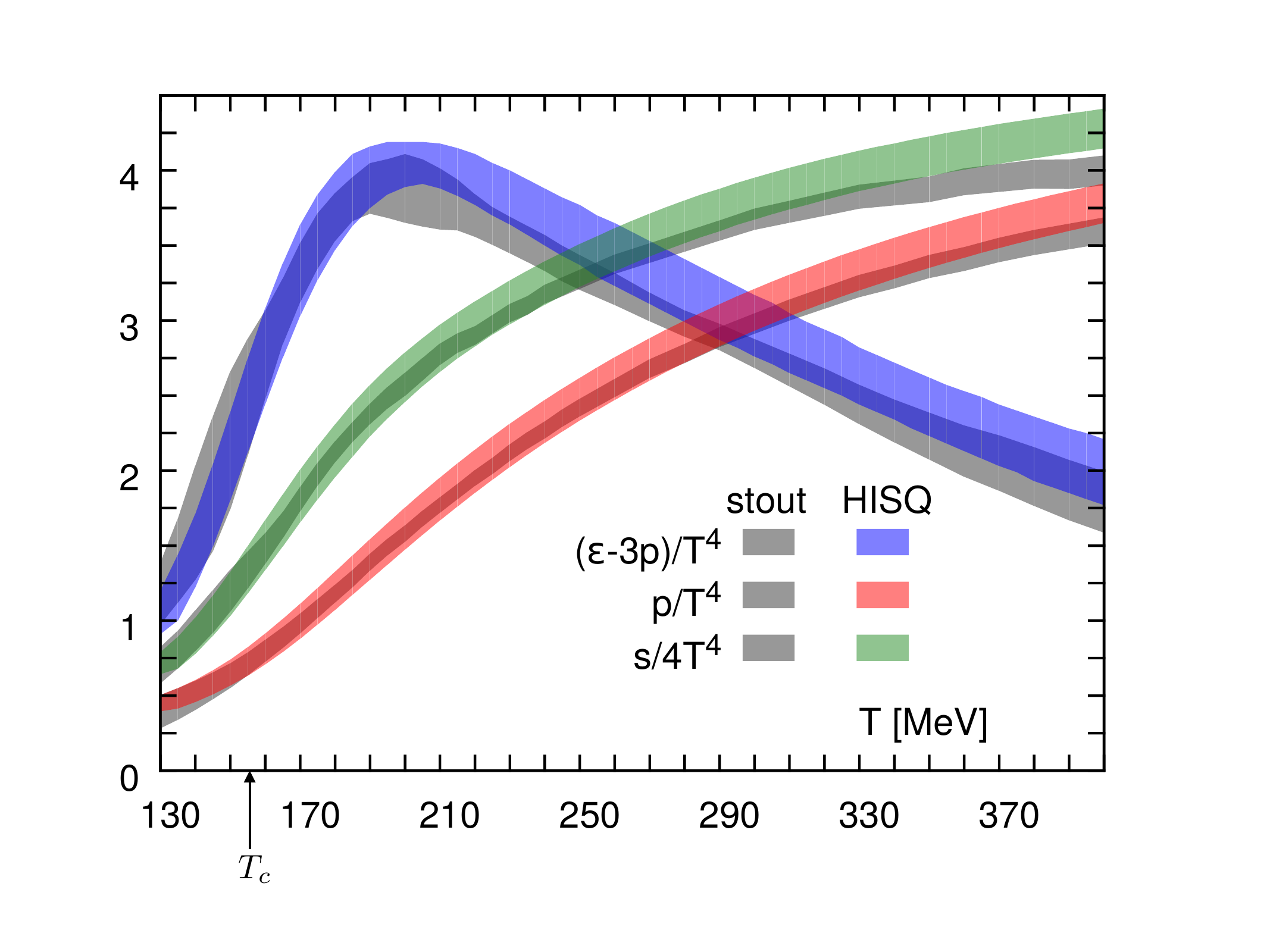}
\caption{$p \beta^4$ as a function of temperature.  (Taken from \cite{Bazavov}).  }
 \label{QCDfig}
\end{figure}

Lattice QCD at finite temperature is well enough developed by now that its data can be used to provide a rough estimate of 
$\rhovac$ based on the proposal of this article.   We refer to the articles \cite{Karsch,Bazavov}.    
Bazavov et. al.  considered 3 flavors with masses $m_u = m_d = m_s/30$ where $m_s \approx 90\, {\rm MeV}$ is 
the strange quark mass,  which is fit to the  $s \bar{s}$ meson mass of $695 \, {\rm MeV}$.    
Data on the pressure is presented in Figure \ref{QCDfig}.    The critical de-confinement temperature is $T_c \approx 155\, {\rm MeV}$.  
In connection with the c-theorem,  from this figure  one sees that $(\rho - 3 p) >0$,   thus $c(r)$ decreases with increasing $r$ based on 
\eqref{ctheoremeq}.

\begin{figure}[t]
\centering\includegraphics[width=.5\textwidth]{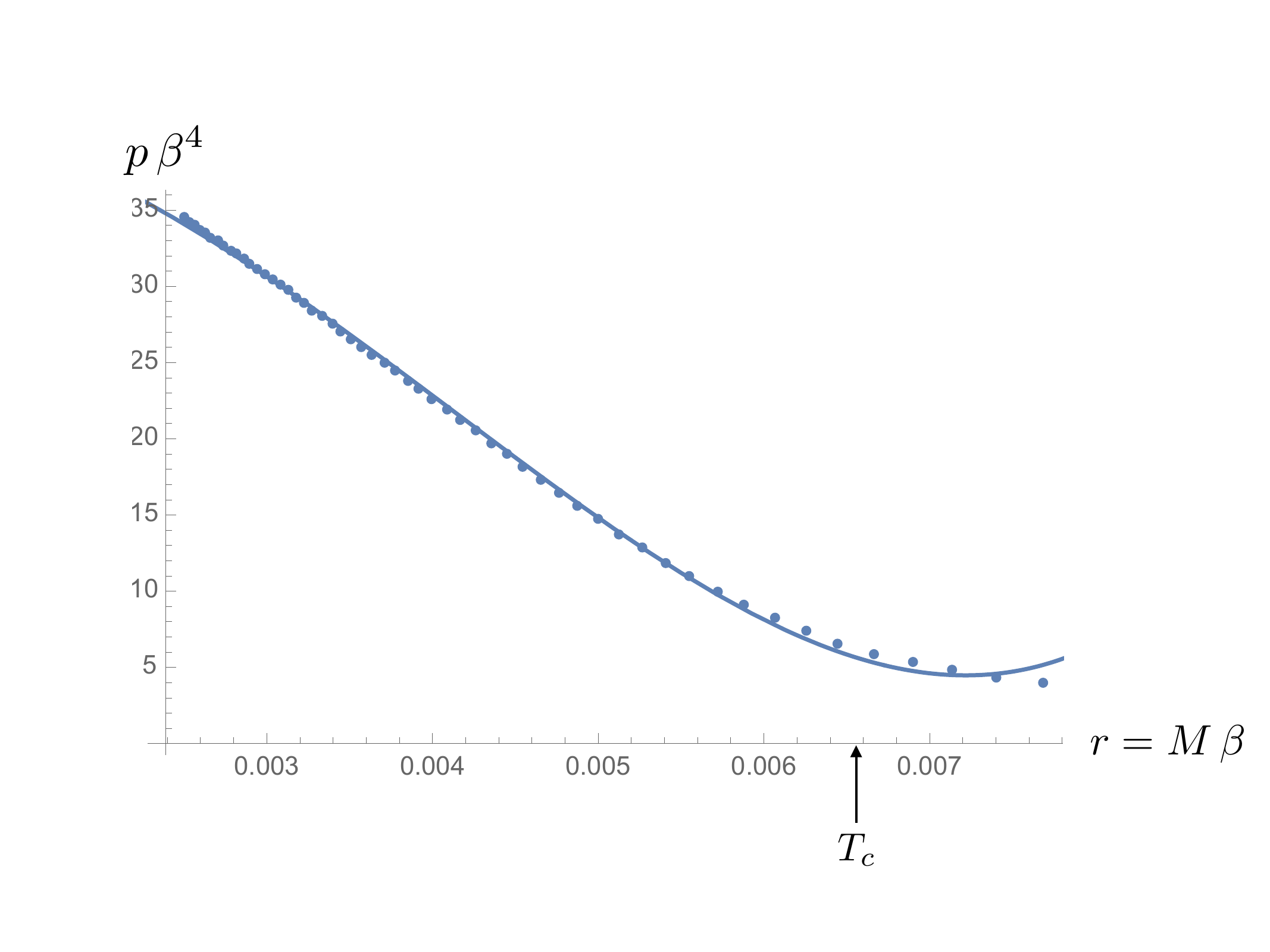}
\caption{The pressure $p \,\beta^4$ as a function of $r=M\beta$ where $M = 1\,{\rm MeV}$.  Data points are from \cite{Bazavov}  which refer to 
Figure \ref{QCDfig},  
and the continuous curve is the fit \eqref{cQCDfit}.} 
 \label{cOfrQCD}
\end{figure}

QCD is UV complete due to asymptotic freedom \cite{Wilczek,Gross}.     We can thus predict $\cuv$.    
Since each of the 8 gluons has 2 polarizations,    and each quark has 3 colors,  2 spins,  and has an anti-quark, for 
$n_f$ flavors  
\beq
\label{cuvQCD}
\cuv = 16 + 21  \, n_f  /2, ~~~~~ (\cuv= 47.5 ~~~~{\rm for} ~~n_f = 3).
\eeq
The data in Figure \ref{QCDfig} can be converted to $c(r) = - \beta^4 \CF/\chi(4) = p \beta^4/\chi(4)$ and is shown in Figure \ref{cOfrQCD}.   
There may be log corrections to the $\cvac \sim r^4$ term as in \eqref{rhovac4D},    however for simplicity with ignore this.   Fitting the data we obtain 
\beq
\label{cQCDfit}
c(r)  \approx 42.72 - 1.468 \cdot 10^6\,  r^2 + c_4  \, r^4,   ~~~~~c_4 = 1.409 \cdot 10^{10} .
\eeq
The fit approximately agrees with the prediction \eqref{cuvQCD},   and presumably is improved with additional higher temperature data.  
In particular,  the above fit gives $\cuv \approx 41.72$ compared to $47.5$ in \eqref{cuvQCD}.  
It's also significant to point out that this fit is to data that is mostly at temperatures above $T_c$,  and one can observe a change of behavior of 
$c(r)$ in the vicinity of $T_c$ where the $\CO(r^4)$  fit in equation \eqref{cQCDfit} begins to fail.

In the Figure \ref{cOfrQCD},  $r = M \beta$ where $M = 1\, {\rm MeV}$.      Thus  \eqref{rhovacC} gives
\beq
\label{rhovacQCD}
\rhovac = - c_4 \,\chi(4) \,  M^4 \approx  - \( 200 \, {\rm MeV} \)^4  \approx - \( 1.3 \, T_c \) ^4  \approx  - \( 2.2 \, m_s \)^4 .
\eeq
Note that $\rhovac$ is {\it negative} and the scale is roughly set by the strange quark mass $m_s$.   Perhaps it is negative since the 
strange quark is a fermion and at least for free theories $\rhovac$ is negative according to \eqref{rhovac4D}.    
It is interesting that $200 \, {\rm MeV}$ is extracted from high temperature data above $T_c$ which a priori does not assume anything about the de-confinement 
phase transition at  the lower temperature.       

\section{Conclusions}

We have presented a non-perturbative definition of the vacuum energy density $\rhovac$ in any spacetime dimension based on Thermodynamics, 
i.e. in the Thermal channel  for a cylinder.      For massive free theories we established modularity in $D$ spacetime dimensions,  
  namely  the equivalence to calculations in a spatially compactified channel \SC.       In 2D,  for integrable QFT's  one can calculate $\rhovac$ exactly with the TBA,   and we presented several 
illustrative examples.    The Thermal channel enjoys certain advantages over the \SC channel since the integrals are not divergent from the beginning, 
and it is easier to identify both pressure $p$  and energy density  $\rho$.      We applied these ideas to QCD and extracted $\rhovac$ for
3 flavor QCD from high temperature lattice data and obtained  a  negative result  $\rhovac \approx - (200 \, {\rm MeV})^4$.    

\bigskip
We close with suggestions for possible further developments.    

\nobull
The TBA in 2D is based solely on the S-matrix and was efficient at calculating the exact $\rhovac$.      It would thus be desirable to work with a formulation of quantum statistical mechanics based on the S-matrix,    since all renormalizations of the theory can be carried out in the usual way at zero temperature and have an S-matrix that is already expressed,  or measured,  in terms of physical masses and couplings.    
There exists such a formulation  in any dimension \cite{Dashen} which leads to the simple formula for the partition function 
\beq
\label{Dashen} 
Z = Z_0 + \inv{2\pi} \int_0^\infty  dE\, \, e^{-\beta E}\, {\rm Tr} \,   \Im \( \d_E \log \hat{S} (E) \).
\eeq
where $\hat{S}$ is the S-matrix quantum operator,  and $Z_0$ is  the free,  unperturbed  partition function.   Unfortunately 
it  turns out to be quite a deal of work to turn the above formula into something useful,  although some results for non-relativistic theories were obtained in \cite{PyeTon,PyeTon2}.     It may be useful to explore this further for relativistic theories in order to perhaps understand if a version of 
the LMB is valid.

\nobull
This work was motivated by the Cosmological Constant Problem which we remarked on in Section II.    We argued that depending on one's assumptions of what the ultimate resolution will be,   approaches to the problem can vary significantly.    For this reason we formulated a well-defined version of the problem where the resolution would amount to properly computing $\rhovac$ in flat Minkowski space,    which is actually  close to its original version \cite{Weinberg}.      Although a big improvement over the naive 120 orders of magnitude error,   for  QCD the  value we obtained is still  too high to explain the measured value \eqref{rhoLambda}.    
Does this rule out our proposal for the CCP?       We wish to argue this is not completely settled.      First of all,   the sign of $\rhovac$ for interacting theories is not easily predictable,  it is negative for QCD,    so there may be cancellations when one considers the complete Standard Model of particle physics.  
This is evident for instance for the massive Thirring model considered above where $\rhovac$ undergoes an infinite number of oscillations around 
zero as one approaches the marginal point   (Figure \ref{SGoscillations}).  
     Secondly,   perhaps there is exists a version of the LMB mechanism in higher dimensions,  wherein all 
S-matrices can be bootstrapped from that of the lightest particle and the $\rhovac$ scale is set by the lightest particle as for integrable 2D theories.     
  This appears challenging to study,   however some promising results were  cultured out of the Swampland \cite{Montero1,Montero2} based on 
  very different ideas in 
  connection with charged black holes.     
   Thirdly,   perhaps one needs to impose the constraint that $\rhovac$ for stable particles are all that contribute to the cosmological constant.   If one could justify the LMB  property to higher dimensions,    this would point to neutrinos to explain the measured $\rhoLambda$ in \eqref{rhoLambda},   and indeed $0.003\, eV$ is close to proposed neutrino masses \cite{neutrino}.   

\nobull
In \cite{LeClairMussardo},  $\langle T_{\mu\nu} \rangle_\beta$  at finite temperature was computed using form-factors in $2D$.     It is worth
exploring if such a form-factor  approach can provide an efficient means to calculate $\rhovac$  in higher dimensions.\footnote{Work along these lines  will be presented elsewere.}

\nobull
We have shown that a restricted modular invariance,  which is a trivial subgroup of  the modular group $SL(2, \ZInteger)$  that interchanges the Thermal and 
\SC channels we referred to as Modularity,  is valid
on the cylinder in $D=2$ and $4$ spacetime dimensions for free massive theories.    It would be very interesting to establish complete 
$SL(2, \ZInteger)$  invariance on the torus,  which has two periodic directions,   for non-conformal theories  in  arbitrary $D$  spacetime dimensions.    Some recent  results concerning this were obtained by Kostov in 2D \cite{Kostov}.   For CFT's in higher dimensions see \cite{Shaghoulian}.


\section{Acknowledgements} 

We wish to  thank  Denis Bernard,  Ivan Kostov,   Peter Lepage,  Giuseppe Mussardo and Matthias Neubert for discussions.

\end{document}